# Windows open for highly tunable magnetostructural phase transitions

Y. Li,[1,2] Z. Y. Wei,[1] H. G. Zhang,[3] E. K. Liu,[1,a)] H. Z. Luo,[2] G. D. Liu,[2] X. K. Xi,[1] S. G. Wang,[1] W. H. Wang,[1] M. Yue,[3] G. H. Wu,[1] and X. X. Zhang[4]

[1] *Beijing National Laboratory for Condensed Matter Physics, Institute of Physics, Chinese Academy of Sciences, Beijing 100190, China*

[2] *School of Materials Science and Engineering, Hebei University of Technology, Tianjin 300130, China*

[3] *College of Materials Science and Engineering, Beijing University of Technology, Beijing 100124, China*

[4] *Division of Physical Science and Engineering, King Abdullah University of Science and Technology, Thuwal 23955-6900, Saudi Arabia*

**Abstract**: An attempt was made to tailor the magnetostructural transitions (MSTs) over a wide temperature range under the principle of isostructural alloying. A series of wide Curie-temperature windows (CTWs) with a maximal width of 377 K between 69 and 446 K were established in the $Mn_{1-y}Co_yNiGe_{1-x}Si_x$ system. Throughout the CTWs, the magnetic-field-induced metamagnetic behavior and giant magnetocaloric effects are obtained. The (Mn,Co)Ni(Ge,Si) system shows great potential as multifunctional phase-transition materials that work in a wide range covering liquid-nitrogen and above water-boiling temperatures. Moreover, general understanding to isostructural alloying and CTWs constructed in (Mn,Co)Ni(Ge,Si) as well as (Mn,Fe)Ni(Ge,Si) are provided.

**Keywords:** Magnetostructural transition, Isostructural alloying, Curie-temperature window, Caloric effect

---

a) Author to whom correspondence should be addressed. E-mail: ekliu@iphy.ac.cn

Magnetic materials with considerable caloric effects in the vicinity of magnetoelastic or magnetostructural transitions (MSTs) show great potential in solid-state refrigeration.[1-3] Currently, a large number of caloric materials have been successively found, including the magnetocaloric materials working at different temperatures.[4] Tunable magnetocaloric effects (MCE) over a wide temperature range may provide expended chance for different applications. Especially, high-temperature MCEs, which may be used at elevated temperatures especially above water boiling point (373 K) for magnetic heat pumps,[5] electric power generation,[6,7] or magnetic cooling,[8,9] are seldom reported.

In recent years, hexagonal MM'X (M, M' = transition metals, X = carbon or boron group elements) compounds have been extensively investigated due to the tunable MSTs and associated giant MCEs.[10-13] In these MM'X compounds, the MSTs, from $Ni_2In$-type hexagonal parent phase to TiNiSi-type orthorhombic martensite with different magnetic states, can be obtained by coupling martensitic structural transitions (MTs) with magnetic transitions. The Curie-temperature windows (CTWs),[12,14] constructed by Curie temperatures of parent ($T_C^A$) and that of martensite ($T_C^M$) phases, were revealed and used to maximize the magnetic-energy change in magnetic-field-induced MSTs. Many MSTs have been tuned into CTWs and exhibited remarkable magnetoresponsive effects.[12-20] During the practice of MST tuning, we further proposed a principle of isostructural alloying to manipulate the structural transitions and the magnetic couplings at one time.[14] The chemically substituting elements were determined from the viewpoint of alloying the isostructural counterparts before performing the experiments. Till now, CTWs have been established in many alloy systems,[21-27] in which the tunable MSTs with giant MCEs were obtained.

In our previously reported results, broad CTWs have been constructed by using isostructural alloying in both $Mn_{1-y}Fe_yNiGe$ and $Mn_{1-y}Co_yNiGe$ systems.[14,21] By further applying isostructural alloying to $Mn_{1-y}Fe_yNiGe$ and MnNiSi, a unprecedentedly wide CTW of 400 K has been achieved.[28] According to the principle of isostructural alloying, a proper isostructural counterpart with a MT and a $T_C^M$ both at high temperatures should be selected to tune MSTs of $Mn_{1-y}Fe_yNiGe$ to high temperatures.[14, 21] This is why MnNiSi was chosen, whose MT temperature ($T_t$) and $T_C^M$ are as high as 1200 K and 622 K respectively.[29-31] In $Mn_{1-y}Co_yNiGe$ system, by doping Co atoms at Mn sites a 230-K CTW from room temperature to 120 K was opened by simultaneously decreasing the MTs and converting the spiral antiferromagnetic (AFM) martensite to ferromagnetic (FM) state. Within the CTW, a giant MCE of -40 J kg$^{-1}$ K$^{-1}$ in a field change of 50 kOe was observed around 236 K for $y = 0.1$. The strongest ferromagnetism of martensite phase was obtained near 120 K for $y = 0.2$. Unfortunately, the desired magnetoresponsive properties disappear for $y > 0.2$ because the MST vanishes below 120 K when it meets $T_C^A$. In MM'X family, there exists an interesting phenomenon that the MT vanishes suddenly when it encounters $T_C^A$,[12, 14, 16, 32] which means that the ferromagnetic ordering at $T_C^A$ will suppress the MTs. For the case in $Mn_{1-y}Co_yNiGe$, it is possible to isostructurally alloy MnNiSi with $Mn_{1-y}Co_yNiGe$, to (i) simultaneously increase $T_t$ and $T_C^M$ and (ii) lower $T_C^A$. For the compositions with higher Co-contents, $T_C^A$ of $Mn_{1-y}Co_yNiGe$ would be decreased to low temperatures as the paramagnetic CoNiGe destroys the FM coupling of MnNiGe parent phase.[33] By doing these, an expanded CTW for tunable MSTs becomes hopeful.

In this work, taking MnNiSi as an isostructural counterpart we create an $Mn_{1-y}Co_yNiGe_{1-x}Si_x$ ($y = 0.2$; $0 \leq x \leq 1$) system by simply substituting Si for Ge in $Mn_{1-y}Co_yNiGe$. The compositions with higher Co-contents ($y = 0.3, 0.4$) are also taken into account. The obtained MSTs can be highly tuned in a wide temperature range from 70 to 450 K. More importantly, in present paper general discussions and summaries on this work and some previous works are made.

Polycrystalline ingots of $Mn_{1-y}Co_yNiGe_{1-x}Si_x$ ($y = 0.2, 0.3, 0.4$; $0 \leq x \leq 1$) alloys were prepared by arc melting high-purity metals four times in argon atmosphere. All ingots were annealed at 1123 K in an evacuated quartz tube for five days and then cooled slowly to room temperature. The phase structures of the samples were characterized by powder x-ray diffraction (XRD) with Cu-$K_\alpha$ radiation. The differential scanning calorimetry (DSC) with permanent-magnet assisted thermogravimetric analysis (TGA) was used to detect the structural and magnetic transitions. With the aid of the upward magnetic pull forces provided by the magnets, a change in sample weight can be detected across the magnetic transition by TGA. The magnetic measurements were performed on superconducting quantum interference device (SQUID) magnetometer, physical property measurement system (PPMS) in the range of 5 ~ 400 K, as well as the vibrating sample magnetometer (VSM, VersaLab, 3 T) for temperatures above 400 K.

Room-temperature XRD analysis of $Mn_{1-y}Co_yNiGe_{1-x}Si_x$ system were performed. Figure 1(a) shows the lattice parameters of $Mn_{0.8}Co_{0.2}NiGe_{1-x}Si_x$ ($0 \leq x \leq 0.45$) samples obtained from XRD patterns (Figure S1 and Table SI in supplemental material).[34] $Mn_{0.8}Co_{0.2}NiGe$ ($x = 0$) shows a $Ni_2In$-type hexagonal structure, which is consistent with our previous study.[21] $Mn_{0.8}Co_{0.2}NiGe_{1-x}Si_x$ series crystallize in a $Ni_2In$-type hexagonal structure for $x \leq 0.25$ and a TiNiSi-type orthorhombic structure for $x \geq 0.30$, which indicates that $T_t$ raises from low temperatures to above the room temperature. With increasing Si content, the cell volumes of both hexagonal and orthorhombic phases decrease due to the smaller size of Si than Ge atom. As shown in Fig. 1(a), large volume expansions of 2.5% and 2.3% during the MT are observed, respectively, for $x = 0.25$ and 0.30, which are consistent with those obtained by temperature-dependent XRD in similar hexagonal alloys.[14, 35]

The M(T) curves of $Mn_{0.8}Co_{0.2}NiGe_{1-x}Si_x$ ($x = 0$, 0.10, 0.15, 0.25 and 0.35) alloy series in a field of 1 kOe between 5 K and 400 K were measured, as shown in Fig. 1(b). For $Mn_{0.8}Co_{0.2}NiGe$ ($x = 0$), the first-order MST from FM parent to FM martensite state, with a clear thermal hysteresis below $T_C^A \sim 129$ K, occurs at 93 K upon cooling (determined as maximal dM/dT). Once Si is introduced on Ge site, MT temperature $T_t$ begins to rise, similar to what is observed in $Mn_{1-y}Fe_yNiGe_{1-x}Si_x$.[28] The tunable first-order MSTs, happening from paramagnetic (PM) hexagonal parent to FM orthorhombic martensite phase, are obtained in alloys with $x$ from 0.10 to 0.35. The highest-temperature MST is observed at 364 K for $x = 0.35$. As shown in Fig. 1(c), the M(T) curves of $Mn_{0.6}Co_{0.4}NiGe_{1-x}Si_x$ ($0.45 \leq x \leq 0.85$) samples in a field of 1 kOe are presented. For $x = 0.45$, no structural transition occurs and only a weak magnetic transition is seen at 69 K. When $x$ is increased to 0.50, the MST recurs suddenly at 93 K, showing a first-order PM-FM transition. For $x \geq 0.50$, $T_t$ increases steadily, spanning over a large temperature range, finally reaches 429 K and $T_C^M$ appears at 446 K when $x = 0.85$.

Based on all data from XRD, magnetic and DSC-TGA measurements (Figs. S1 and S2 in supplemental material),[34] we propose a magnetostructural phase diagram (see complete phase diagram in Fig. S3)[34] of $Mn_{1-y}Co_yNiGe_{1-x}Si_x$ ($y = 0.2, 0.3, 0.4$; $0 \leq x \leq 1$), as shown in Fig. 2(a). For all alloy series, the transition temperature $T_t$ can be raised to high temperatures by Si substitution and encounters $T_C^M$ (also upper critical temperature, $T_{cr}$) at high temperatures of 409 K, 429 K and 446 K for $y = 0.2, 0.3, 0.4$, respectively. With increasing Co content ($y$), at the same time $T_C^A$ (also lower critical temperature, $T_{cr}$) of CTW decreases from 125 K to 69 K. These behaviors are consistent with those in $Mn_{1-y}Fe_yNiGe_{1-x}Si_x$ system.[28] This brings about an encouraging result that the width of the CTW between upper and lower $T_{cr}$s is increased from 230 K to 377 K. The broad and extended CTW series are realized in $Mn_{1-y}Co_yNiGe_{1-x}Si_x$ system.

According to our previous study,[21] high Co content can highly stabilize the parent phase of $Mn_{1-y}Co_yNiGe$ since the isostructural counterpart CoNiGe has a stable parent structure, as shown in Fig. 2(b). In principle, higher Co content would bring the MT to very low temperatures, even below the absolute temperature if Co content is high enough. In order to gain a complete and deep understanding on the magnetostructural transitions in alloy series with higher Co contents ($y > 0.2$), we extrapolated the real transition temperatures above $A_0$ point ($T_t \sim 93$ K of $Mn_{0.8}Co_{0.2}NiGe$ compound) to get the theoretical martensitic transition temperature $T_t$, as shown by the broken line. For the studied $y = 0.3$ and $0.4$, the theoretical $T_t$ reaches at $B_0$ and $C_0$ points, where the parent phase has been deeply stabilized. Owing to high $T_t$ of MnNiSi, introducing Si can weaken the stability of parent phase and immediately raise $T_t$ of $Mn_{1-y}Co_yNiGe_{1-x}Si_x$, as demonstrated by the case of $y = 0.2$ alloy series in Fig. 2(a) and (b). The theoretical $T_t$ of $y = 0.3$ and $0.4$ alloy series will also increase from $B_0$ and $C_0$, respectively, to higher temperatures. With parent phase being stabilized deeply, nevertheless, it needs more Si contents ($x = 0.26$ for $y = 0.3$ and $x = 0.48$ for $y = 0.4$) to awaken the "dead" transition above $T_C^A$ (the lower $T_{cr}$, denoted by $A_1$, $B_1$ and $C_1$ in Fig. 2(a)). As mentioned above, for MM'X compounds the magnetic ordering around $T_C^A$ can suppress the MTs.[12, 14, 16, 32] Higher Co content results in a decrease in $T_C^A$ from 125 K to 69 K, due to the paramagnetism of isostructural CoNiGe.[33] The transitions can be thus observed at lower temperatures where the suppression is removed. In this case, Co substitution at Mn sites provides a larger temperature space for Si substitution to raise the MST (for example, from $B_1$ to $B_2$). For high Si contents, $T_C^M$ (the upper $T_{cr}$) of the system is increased from 409 K ($A_2$) to 446 K ($C_2$) because of the high $T_C^M$ of MnNiSi counterpart. The combined effects of Co- and Si- substitution from the view of isostructural alloying thus lead to a series of broad CTWs for highly tunable MSTs.

From the results, one can see that for MnNiX based compounds introducing Co at Mn sites plays an essential role in keeping $T_C^A$ (the lower $T_{cr}$) of the studied systems at low temperatures. This situation is also applicable to $Mn_{1-y}Fe_yNiGe$,[14] $Mn_{1-y}Fe_yNiGe_{1-x}Si_x$[28] and $Mn_{0.5}Fe_{0.5}NiSi_{1-x}Al_x$[23] systems in which Fe atoms were also introduced to Mn sites. Only in this case, can the CTW that covers low temperatures be established and can the MSTs be tuned to low temperatures without suppression by the magnetic ordering of parent phase. Meanwhile, Co or Fe at Mn sites plays a key role in keeping the strong FM coupling in martensite, which is quite important in MnNiX based compounds. In contrast, due to the high $T_C^A$ of isostructural counterparts MnCoGe (260 K) and MnFeGe (228 K), introducing Co or Fe at Ni sites always makes $T_C^A$ of the studied systems such as $MnNi_{1-x}Fe_xGe$ very high,[25, 28, 32] which narrows the CTWs and constrains the MSTs at high temperature range. Nowadays, almost all the MSTs of MM'X compounds are studied within the CTWs, in order to get large magnetization changes and desired properties. During the composition design, it is necessary to pay more attention to this difference of atom occupying sites and the atom-resolved magnetic exchange interactions, which is an important base for accurate applications of the so-called isostructural alloying principle.[14, 21, 28] Confusion of site difference and negligence of CTW will lead to inefficient results or even wrong conclusions, like the case in Ref. 36. Here it should be further stated that, if the alloys are in equilibrium state Co (Fe) atoms prefer the Mn sites than Ni sites in $Mn_{1-y}Co_yNiGe_{1-x}Si_x$ ($Mn_{1-y}Fe_yNiGe_{1-x}Si_x$) system, which is dominated by the atom site occupation (valence electron) rule of MM'X compounds.[14, 21, 37, 38]

In order to further analyze the ferromagnetism behavior within the broad CTWs obtained in this study, we measured the magnetization M(H) curves of $Mn_{1-y}Co_yNiGe_{1-x}Si_x$ samples ($y$ = 0.2, 0.3, 0.4) at 5 K. The related $M_S$ and $H_S$ are shown in Fig. 3. For $Mn_{1-y}Co_yNiGe$ series ($x$ = 0),[21, 32] with increasing Co content the martensite phase changes from the spiral AFM state with a high saturation field ($H_s$) of about 100 kOe to a FM state with a relatively low $H_s$ of about 2.9 kOe, as shown in Inset I of Fig. 3. For $Mn_{0.8}Co_{0.2}NiGe_{1-x}Si_x$ ($y$ = 0.2) series, after introducing Si to the system the $H_s$ further decreases from 2.9 kOe ($x$ = 0) to 1.74 kOe ($x$ = 0.45). Inset II of Fig. 3 depicts the magnetization curve at 5 K of an example alloy with $y$ = 0.2 and $x$ = 0.20, showing a typical FM behavior with a low $H_s$, which is desired for the low-field effects. At the same time, $M_s$ of this alloy series keeps basically unchanged with high values of 70 ~ 80 emu g$^{-1}$, which benefits the large magnetization change ($\Delta M$) across MSTs. For higher Co content of $y$ = 0.3 and 0.4, similar trends are observed. Stable values of $M_s$ are maintained during the Si substitution, which can be attributed to the slight influence of the main-group elements in Ni-(Ge,Si) covalent networks on magnetic moments of metallic Mn/Co atoms.[37, 39] Within the CTWs, the various alloy series, exhibiting high $M_s$ and low $H_s$, are expected to enhance the magnetoresponsive effects including large MCEs.

The entropy changes ($\Delta S$) of the studied alloys were derived using Maxwell relation,[1] $\Delta S = \int_0^H (\frac{\partial M}{\partial T})_H dH$, from the isothermal magnetization curves across the MSTs. In order to avoid overestimating the $\Delta S$ of MSTs with a thermal hysteresis, the isothermal magnetization curves of $x$ = 0.10 and 0.25 in $Mn_{0.8}Co_{0.2}NiGe_{1-x}Si_x$ series, shown in Fig. 4(a) and 4(b), were measured using the temperature loop method.[40] All samples exhibit a negative (conventional) $\Delta S$ because the martensite is FM and the austenite is PM. Figure 4(c) and 4(d) shows the MCEs of $x$ = 0.10 and 0.25 in $Mn_{0.8}Co_{0.2}NiGe_{1-x}Si_x$ series. For both samples, giant MCEs with -35 and -40 J kg$^{-1}$ K$^{-1}$ at a field change of 50 kOe are obtained. In particular, the MCEs at a low field of 20 kOe are also very large with values up to -19 and -16 J kg$^{-1}$ K$^{-1}$. These values are comparable with or even larger than many other MCE materials.[15, 16, 18, 21] The (low-field) giant MCEs in the CTWs are mainly attributed to the low $H_s$ and high $M_s$ of the FM martensite phase (Fig. 3). Similarly, Figure 4(e) and 4(f) show the MCEs of samples with $x$ = 0.65 in $Mn_{0.7}Co_{0.3}NiGe_{1-x}Si_x$ ($y$ = 0.3, $x$ = 0.65) and $x$ = 0.70 in $Mn_{0.6}Co_{0.4}NiGe_{1-x}Si_x$ ($y$ = 0.4, $x$ = 0.70) series. Large MCEs of -8 and -10 J kg$^{-1}$ K$^{-1}$ at a field change of 30 kOe are respectively obtained. For the sample ($y$ = 0.3, $x$ = 0.65), in particular, the large, high-temperature MCE happens around 445 K. This temperature is much higher than those of many giant MCE materials. The large MCEs based on the highly tunable MSTs in the wide temperature range can be expected to provide a potential for wide-temperature-range applications, including the high-temperature magnetic cooling, magnetic heat pumps, and electric power generation above 400 K.

A remarkable characteristic of MM'X alloys is their large volume expansion upon MSTs. As shown in Fig. 5, $Mn_{1-y}Co_yNiGe_{1-x}Si_x$, combined with $Mn_{0.965}CoGe$, $MnNi_{0.8}Fe_{0.2}Ge$ and $Mn_{1-y}Fe_yNiGe_{1-x}Si_x$,[12, 14, 28] shows large volume expansion ($\Delta V/V$) with 2.2% to 4% during MSTs. In other words, Si substitution almost does not reduce the large volume change during MSTs. The uniaxial strains along $c_h$ ($a_o$) reach 11.2% for $Mn_{1-y}Co_yNiGe_{1-x}Si_x$ (Fig. 1(a)) and 12.4% for $Mn_{1-y}Fe_yNiGe_{1-x}Si_x$.[28] The large $\Delta V/V$ may enhance the MCEs originated from magnetic and structural entropy changes.[41] At the same time, these large $\Delta V/V$ and uniaxial strain may bring considerable mechanocaloric effects (mCEs) including barocaloric and elastocaloric effects.[3, 42, 43, 44, 45, 46] Figure 6 shows the latent heats ($\Delta E$) during MTs for the $Mn_{1-y}Co_yNiGe_{1-x}Si_x$ and $Mn_{1-y}Fe_yNiGe_{1-x}Si_x$ systems, which are rather large among the materials with first order transitions. The large $\Delta E$ and the relatively small heat capacity ($C_p$, around 1.4 J g$^{-1}$ K$^{-1}$)[25] of this kind of material will lead to large adiabatic temperature change associated with MSTs. For the MST from PM parent phase to FM martensite phase in the present system, the two entropy changes enhance the total entropy change because they are both negative.

In summary, we have established a series of wide Curie-temperature windows that cover the temperature range from the cryogenic (70 K), through the room (300 K) and to the high temperatures (450 K) in single-phase $Mn_{1-y}Co_yNiGe_{1-x}Si_x$ system, by simultaneously upraising $T_t$ and $T_C^M$ of the system under the principle of isostructural alloying. The first-order magnetostructural transitions can be highly tuned in the quite large temperature range. The giant entropy changes as large as -40 J kg$^{-1}$ K$^{-1}$ at 50-kOe field change are observed at 295 K. In particular, a high-temperature large entropy changes of -8 J kg$^{-1}$ K$^{-1}$ at 30-kOe field change is gained near 450 K. Based on these magnetostructural transitions in the windows, the large MCEs, high-temperature MCEs, high tunability, large volume expansion and uniaxial strain jointly make the (Mn,Co)Ni(Ge,Si) system is of great potential for solid state cooling, heat pumps, functional gradient materials, energy conversion in different temperature regions. The principle of isostructural alloying and Curie-temperature window have shown the effectivity as guideline in tuning structural and magnetic transitions and in designing magnetostructural materials.

*Note added:* During the review process of this paper, we became aware of a related study in a case of $y = x$ in $Mn_{1-y}Co_yNiGe_{1-x}Si_x$ (J. Liu, et al., Sci. Rep. **6**, 23386 (2016)), providing more understanding to the whole phase diagram established in our work.

**ACKNOWLEDGEMENTS**

This work was supported by National Natural Science Foundation of China (51301195, 51401002, and 51431009), Beijing Municipal Science & Technology Commission (Z141100004214004), and Youth Innovation Promotion Association of Chinese Academy of Sciences (2013002).

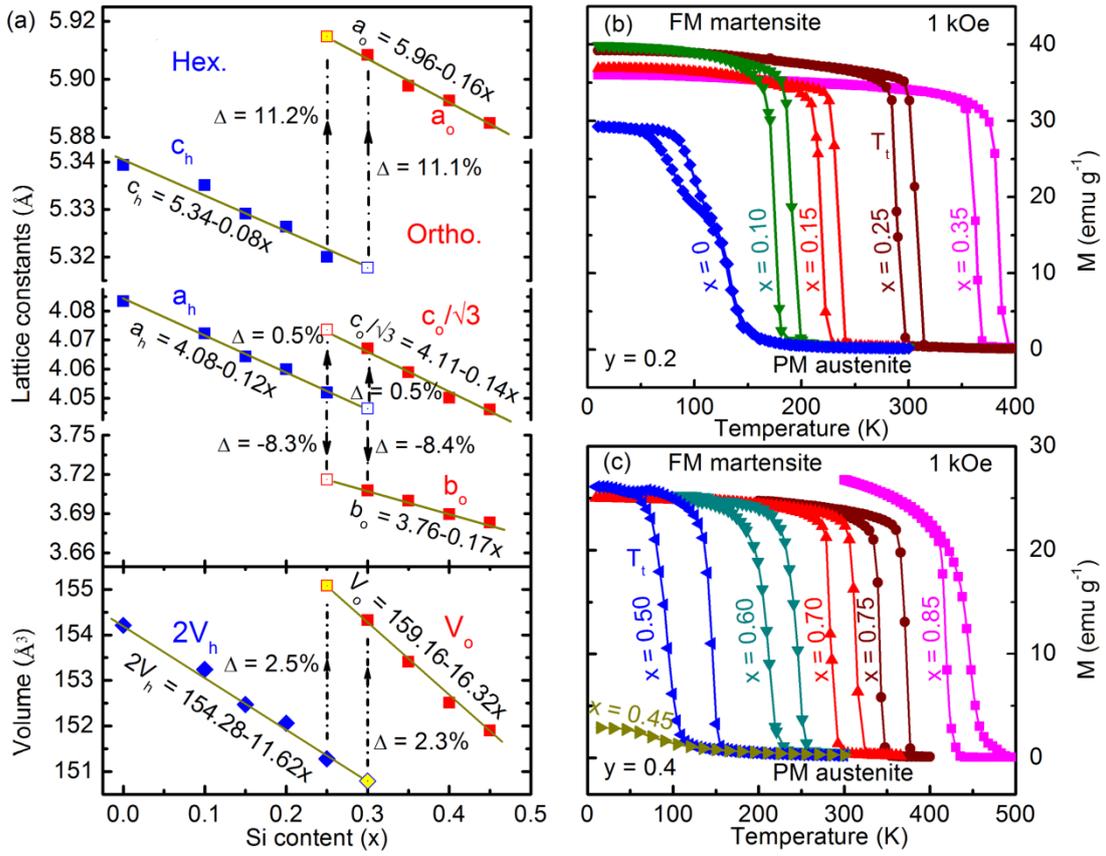

FIG. 1. Si-content dependence of (a) room-temperature XRD patterns and cell volumes of hexagonal and orthorhombic phases of $Mn_{0.8}Co_{0.2}NiGe_{1-x}Si_x$. There is a relation between volumes of hexagonal ($V_h$) and orthorhombic ($V_o$) phases: $V_o \sim 2V_h$, $\Delta V/V_h = (V_o-2V_h)/2V_h$.[29] (b) Temperature dependence of the magnetization (M(T) curves) in the field of 1 kOe of $Mn_{0.8}Co_{0.2}NiGe_{1-x}Si_x$. (c) M(T) curves in the field of 1 kOe of $Mn_{0.6}Co_{0.4}NiGe_{1-x}Si_x$.

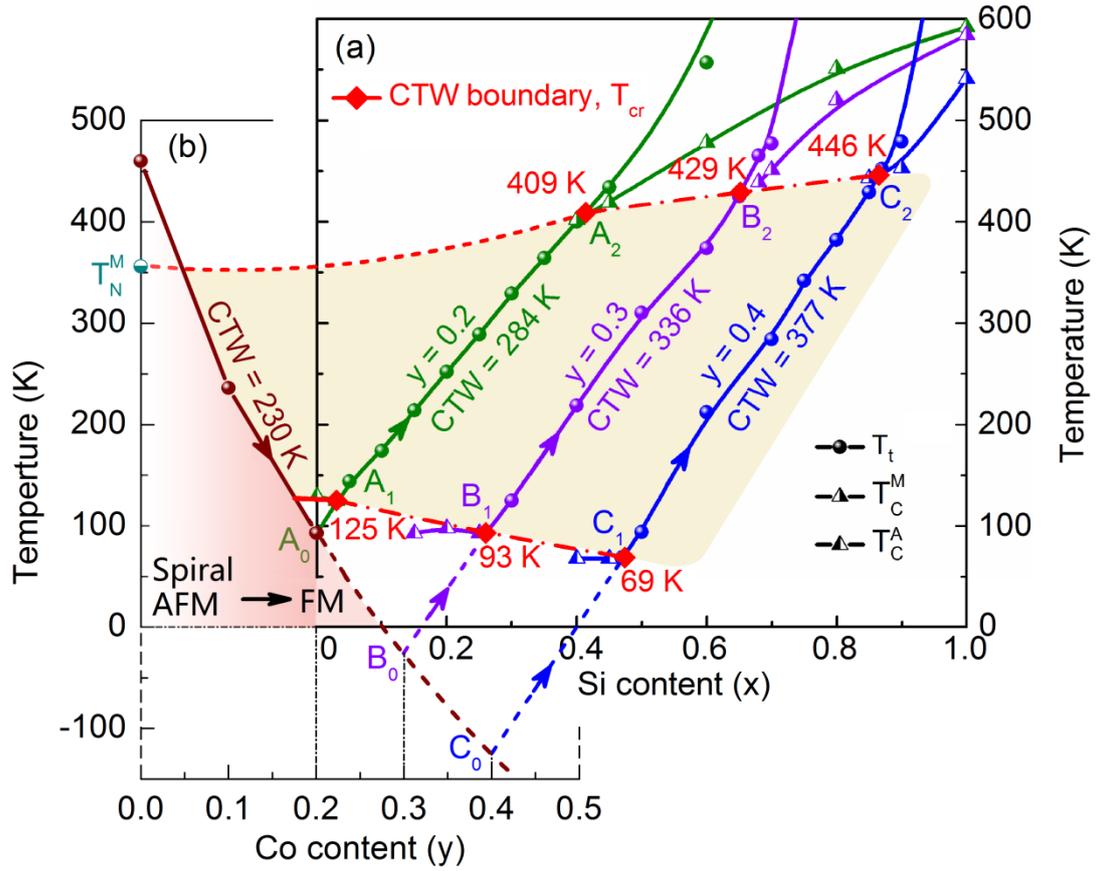

FIG. 2. Magnetostructural phase diagram of (a) $Mn_{1-y}Co_yNiGe_{1-x}Si_x$ ($y$ = 0.2, 0.3, 0.4; $0 \leq x \leq 1$) and (b) $Mn_{1-y}Co_yNiGe$ ($0 \leq y \leq 0.2$) system with wide Curie-temperature windows (CTWs), by introducing the Si and Co elements based on the isostructural alloying principle, respectively. The light yellow region demonstrates the region of CTWs between $T_C^M$ and $T_C^A$.

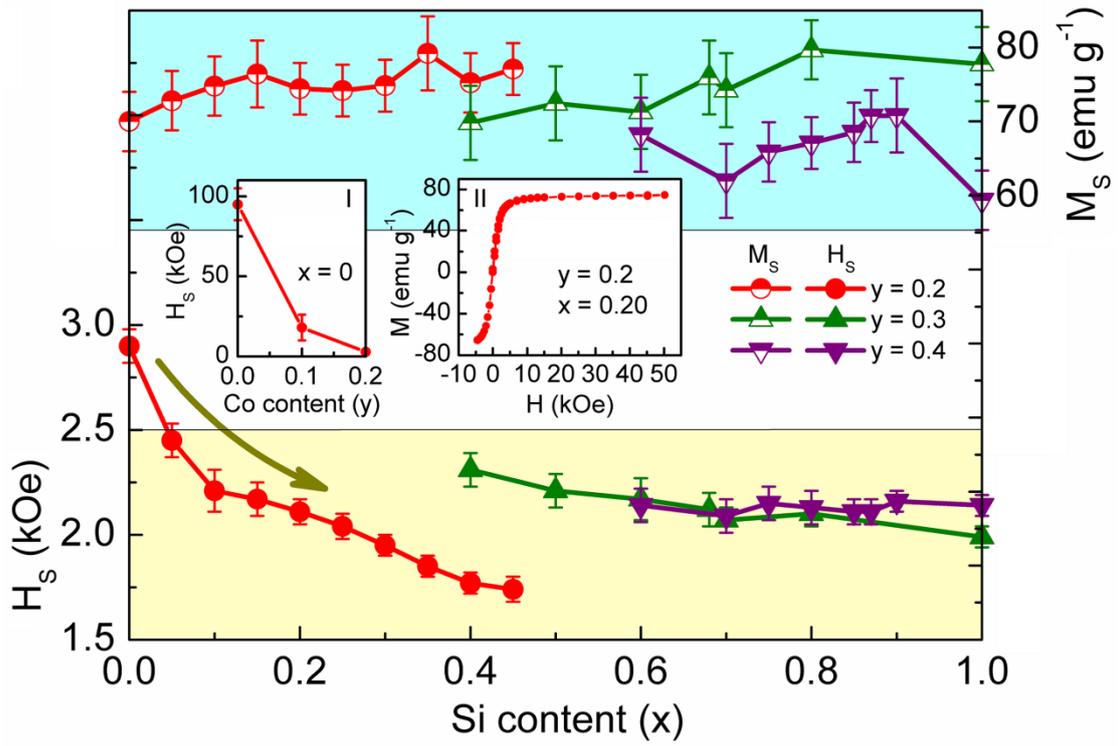

FIG. 3. Saturation field ($H_S$) and saturation magnetization ($M_S$) of $Mn_{1-y}Co_yNiGe_{1-x}Si_x$ ($y$ = 0.2, 0.3, 0.4; $0 \leq x \leq 1$) samples at 5 K. Inset I shows the Co-content dependence of $H_S$ for $Mn_{1-y}Co_yNiGe$ (Data were taken from refs. 21 and 31). Inset II shows the magnetization curve of example alloy of $Mn_{0.8}Co_{0.2}NiGe_{0.80}Si_{0.20}$.

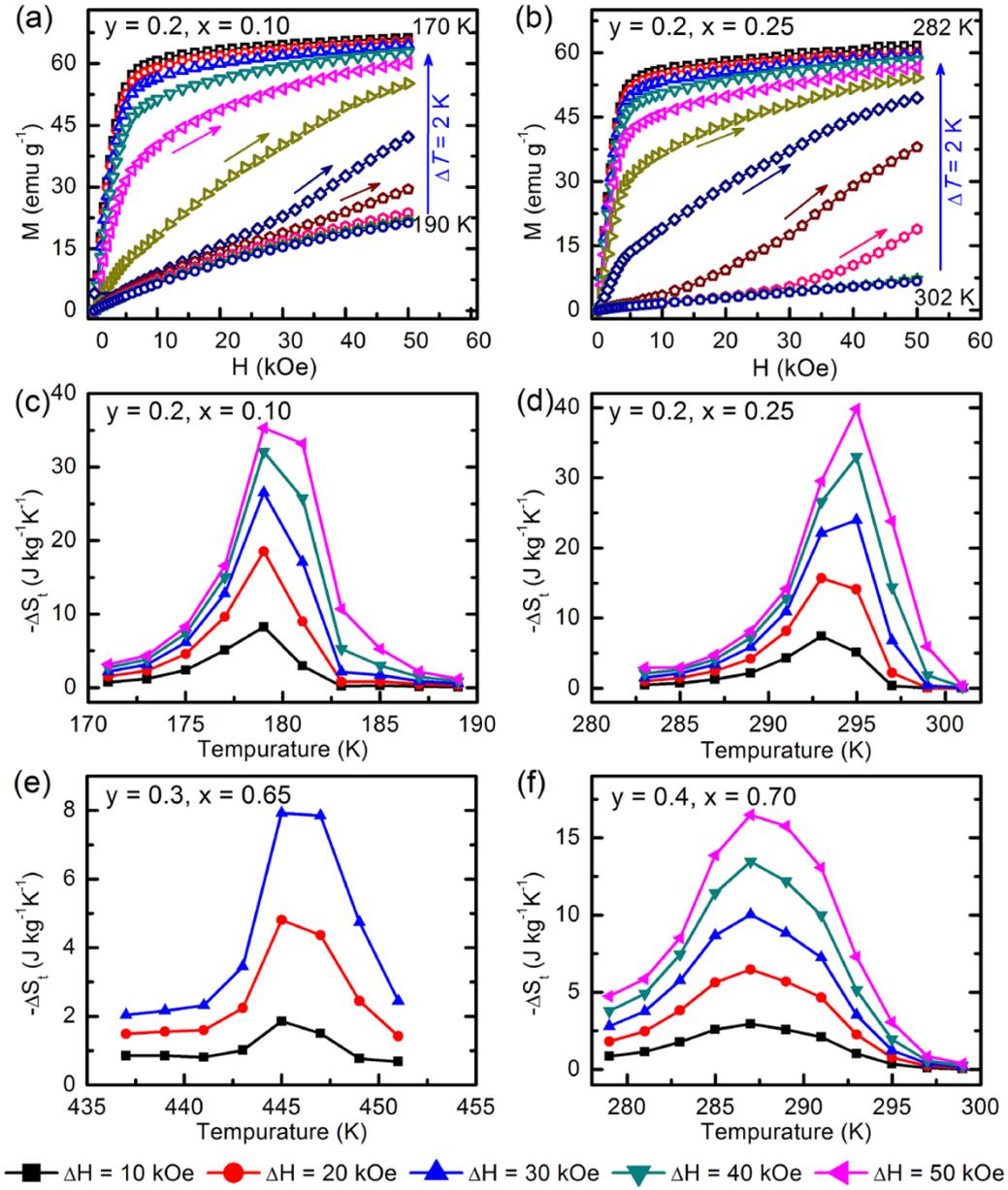

FIG. 4. Isothermal magnetization curves as a function of magnetic field at various temperatures of (a) ($y = 0.2$, $x = 0.10$), (b) ($y = 0.2$, $x = 0.25$). Isothermal entropy changes (ΔS) in various field changes of (c) ($y = 0.2$, $x = 0.10$), (d) ($y = 0.2$, $x = 0.25$), (e) ($y = 0.3$, $x = 0.65$), (f) ($y = 0.4$, $x = 0.70$).

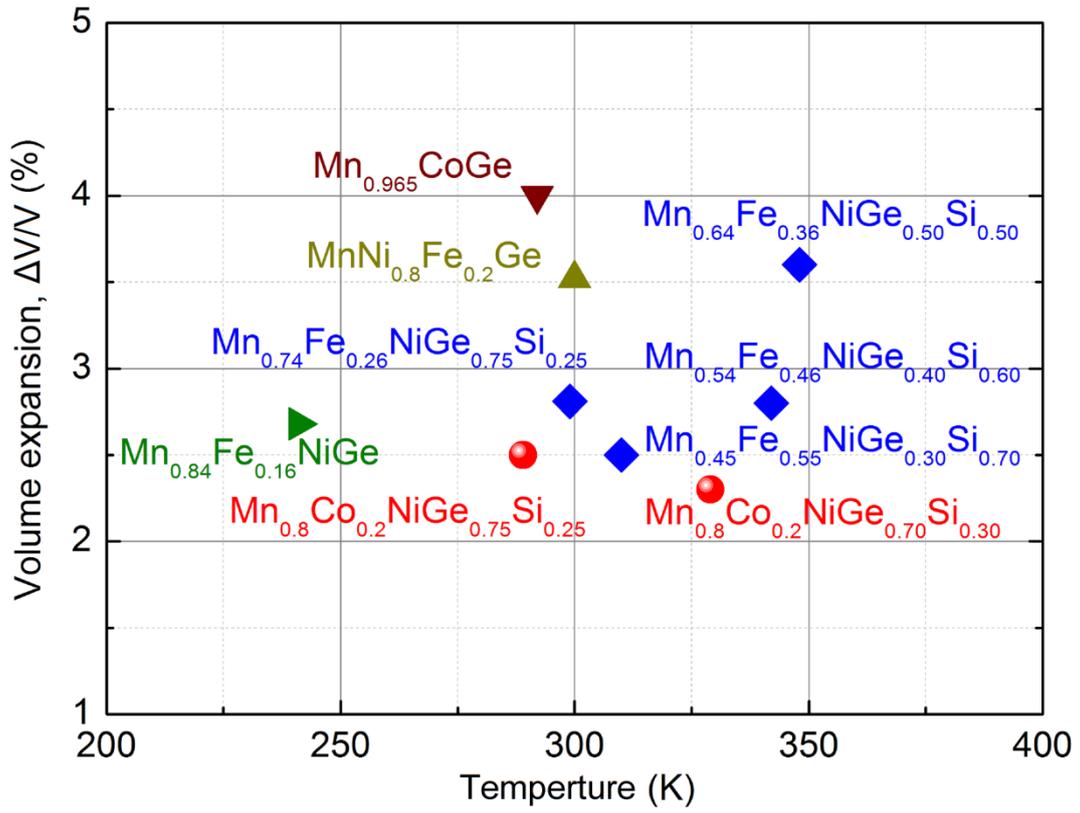

FIG. 5. Volume expansion ($\Delta V/V$) of $Mn_{1-y}Co_yNiGe_{1-x}Si_x$, $Mn_{1-y}Fe_yNiGe_{1-x}Si_x$[28] and some typical MM'X alloys with MSTs.[12, 14]

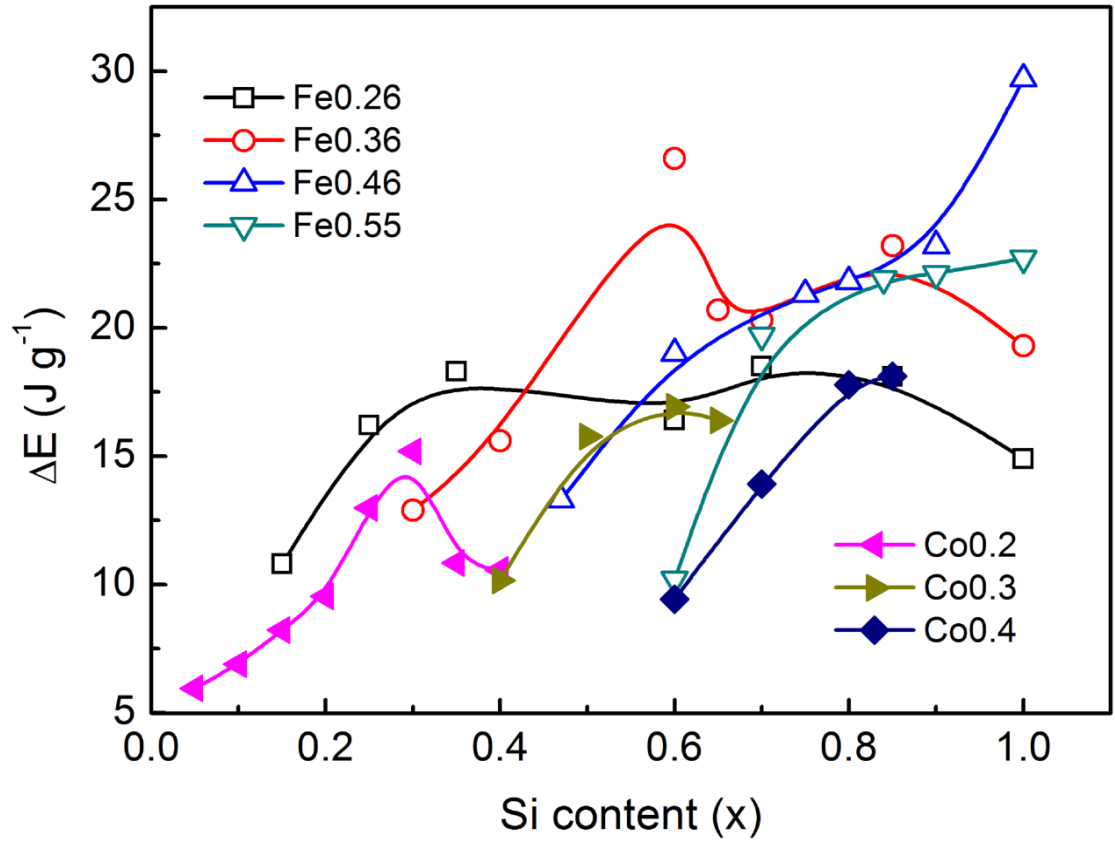

FIG. 6. Latent heat (ΔE) during MSTs of $Mn_{1-y}Co_yNiGe_{1-x}Si_x$ and $Mn_{1-y}Fe_yNiGe_{1-x}Si_x$[28] systems.

# Supplemental Material

# Windows open for highly tunable magnetostructural phase transitions

Y. Li,[1,2] Z. Y. Wei,[1] H. G. Zhang,[3] E. K. Liu,[1,a)] H. Z. Luo,[2] G. D. Liu,[2] X. K. Xi,[1] S. G. Wang,[1] W. H. Wang,[1] M. Yue,[3] G. H. Wu,[1] and X. X. Zhang[4]

[1] *Beijing National Laboratory for Condensed Matter Physics, Institute of Physics, Chinese Academy of Sciences, Beijing 100190, China*

[2] *School of Materials Science and Engineering, Hebei University of Technology, Tianjin 300130, China*

[3] *College of Materials Science and Engineering, Beijing University of Technology, Beijing 100124, China*

[4] *Division of Physical Science and Engineering, King Abdullah University of Science and Technology, Thuwal 23955-6900, Saudi Arabia*

**Contents:**

I. Room-temperature XRD analysis of $Mn_{0.8}Co_{0.2}NiGe_{1-x}Si_x$;

1) Room-temperature XRD patterns of $Mn_{0.8}Co_{0.2}NiGe_{1-x}Si_x$;

2) Room-temperature phase structures, lattice parameters(Å) of $Mn_{0.8}Co_{0.2}NiGe_{1-x}Si_x$;

II. DSC-TGA analysis for $Mn_{0.8}Co_{0.2}NiGe_{1-x}Si_x$;

III. The complete phase diagram of $Mn_{1-y}Co_yNiGe_{1-x}Si_x$.

---

a) Author to whom correspondence should be addressed. E-mail: ekliu@iphy.ac.cn

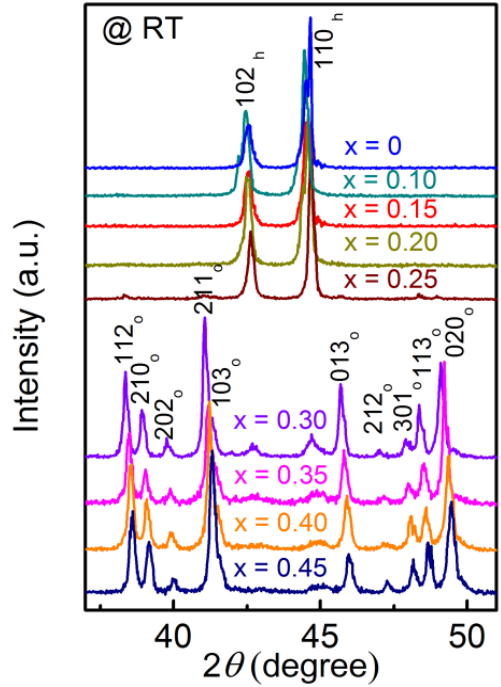

**Figure S1. Room-temperature XRD patterns of $Mn_{0.8}Co_{0.2}NiGe_{1-x}Si_x$.**

At room temperature, the samples crystallize in a $Ni_2In$-type hexagonal parent structure when $x \leq 0.25$ and a TiNiSi-type orthorhombic martensite structure for $x \geq 0.30$. The diffraction peaks of parent and martensite phases are indexed.

**TABLE SI. Room-temperature phase structures, lattice parameters (in Å) of $Mn_{0.8}Co_{0.2}NiGe_{1-x}Si_x$ ($y = 0.2$, $0 \leq x \leq 0.45$).**

| $x$ | Structure | $a_o$ ($c_h$) (Å) | $b_o$ ($a_h$) (Å) | $c_o$ ($\sqrt{3}a_h$) (Å) | $V_o$($2V_h$) (Å$^3$) | $(V_o-V_h)/V_h$ (%) |
|---|---|---|---|---|---|---|
| 0 | Hex. | 5.3394 | 4.0834 | 7.0727 | 154.21 | |
| 0.10 | Hex. | 5.3351 | 4.0723 | 7.0534 | 153.24 | |
| 0.15 | Hex. | 5.3291 | 4.0643 | 7.0396 | 152.47 | |
| 0.20 | Hex. | 5.3264 | 4.0599 | 7.0320 | 152.06 | |
| 0.25 | Hex. | 5.3200 | 4.0519 | 7.0181 | 151.28 | ~2.5 |
| | Ortho. | 5.9148 | 3.7161 | 7.0556 | 155.08† | |
| 0.30 | Hex. | 5.3177 | 4.0464 | 7.0086 | 150.79‡ | ~2.3 |
| | Ortho. | 5.9083 | 3.7077 | 7.0444 | 154.32 | |
| 0.35 | Ortho. | 5.8977 | 3.6999 | 7.0303 | 153.41 | |
| 0.40 | Ortho. | 5.8927 | 3.6895 | 7.0149 | 152.51 | |
| 0.45 | Ortho. | 5.8849 | 3.6833 | 7.0080 | 151.90 | |

The lattice parameters are given in an orthorhombic description in order to compare the changes of parameters. The red data (for martensite of $x = 0.25$ and parent phase of $x = 0.30$) are extrapolated by fitting.

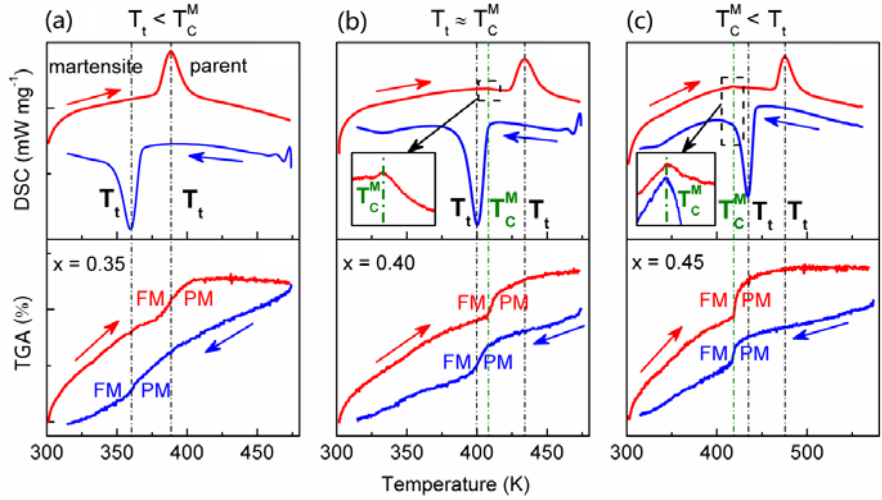

**Figure S2.** DSC-TGA analysis on martensitic structural and Curie magnetic transitions at high temperatures for $Mn_{0.8}Co_{0.2}NiGe_{1-x}Si_x$, $x = 0.35$ (a), 0.40 (b) and 0.45 (c). Insets to (b) and (c) show the DSC peaks of the $T_C^M$.

All the MSTs and Curie magnetic transitions of $Mn_{1-y}Co_yNiGe_{1-x}Si_x$ samples at high temperatures above 400 K were further measured by DSC-TGA. Figure S2 presents the results of $Mn_{0.8}Co_{0.2}NiGe_{1-x}Si_x$ sample series. All transition temperatures are determined by the peak maximum in DSC curves. Consistent with the case in magnetic measurements below 400 K (Figure 1(b) in main context), $T_t$ continuously increases with increasing Si content ($x$). For $x = 0.35$ (Figure S2(a)), the first-order structural MT occurs with a remarkable enthalpy change. Synchronously, a measured change of sample weight was detected in both heating and cooling processes, as shown in TGA curves in Figure S2(a). This indicates that the sample changes its structure between PM parent and FM martensite phases, which is coherent with the results characterized by M(T) curves in Figure 1(b) in main context. From the results, one can see the MST of $x = 0.35$ still locates in the CTW ($T_t < T_C^M$). For $x = 0.40$ (Figure S2(b)), in the cooling process a similar weight change was observed across the forward transition. However, upon heating there is no change in TGA curve across the reverse transition. Instead, a remarkable weight change was observed between the forward and reverse transitions. An enlarged image of DSC curve (inset to Figure S2(b)) reveals the magnetic transition at Curie temperature ($T_C^M \sim 408$ K) of martensite, which indicates $T_C^M$ is also raised by Si substitution from low temperature (~ 350 K). For this sample, its MST just happens at the upper critical temperature ($T_{cr}$) of CTW, that is $T_t \approx T_C^M$ in this sample. Even upon heating the MST decouples and the structural MT happens without a magnetic state change. One can further see that the MTs in samples with $x > 0.40$ will step out of the CTW. This case is observed in the sample with $x = 0.45$, as shown in Figure S2(c). Both $T_t$s in heating and cooling processes go to higher temperatures than $T_C^M \sim 419$ K ($T_t > T_C^M$). From the DSC-TGA analysis, it can be seen that $T_t$ and $T_C^M$ are both increased and they meet together at about 409 K in $Mn_{0.8}Co_{0.2}NiGe_{1-x}Si_x$ sample series. A width of CTW as 280 K is obtained between $T_C^A \sim 125$ K and $T_C^M \sim 409$ K in $Mn_{0.8}Co_{0.2}NiGe_{1-x}Si_x$ sample series. Similar measurements and analysis at high temperatures have been performed for the alloys with $y = 0.3$ and 0.4.

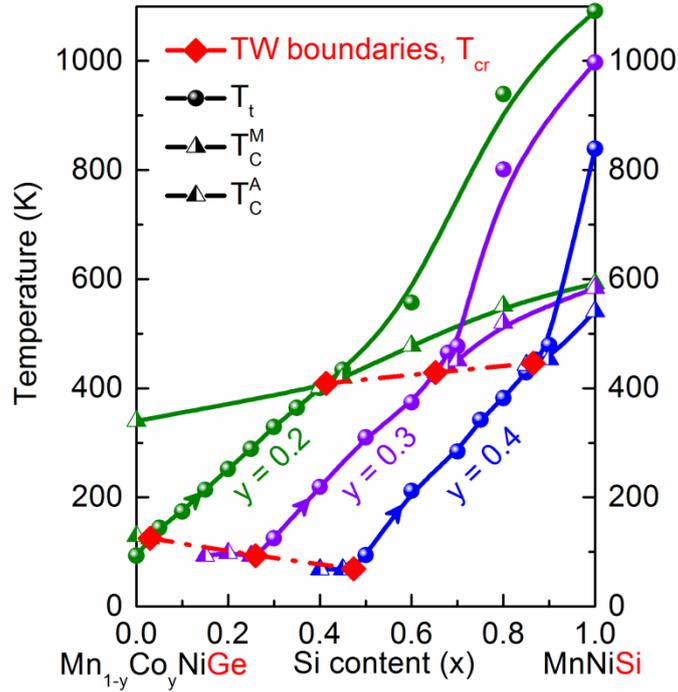

**Figure S3. The complete phase diagram of $Mn_{1-y}Co_yNiGe_{1-x}Si_x$ ($y$ = 0.2, 0.3, 0.4; $0 \leq x \leq 1$).**

Based on all data from XRD, magnetic and DSC-TGA measurements, we provide a complete phase diagram of $Mn_{1-y}Co_yNiGe_{1-x}Si_x$ ($y$ = 0.2, 0.3, 0.4; $0 \leq x \leq 1$), as shown in Figure S3. As we can clearly see, the $T_t$ increases with increasing Si content for $y$ = 0.2, 0.3, 0.4. In case of low Si contents, $T_t < T_C^A$, no MT can be observed due to the suppression of magnetic ordering at $T_C^A$. With increasing Si content, $T_t$ happens from paramagnetic parent to ferromagnetic martensite phase between the two red dashed lines, which means that magnetic and structural transitions are coupled together. Further increasing Si content, $T_t$ is higher than $T_C^M$ and the MT happens at high-temperature paramagnetic state.